\documentclass{ifacconf}

\usepackage{multirow}
\usepackage{graphicx}      
\usepackage{natbib}        
\usepackage[left=0.6in,right=0.5565in,bottom=0.53in,top=1.45in]{geometry}

\usepackage[T1]{fontenc}
\usepackage{lmodern}

\usepackage{amsmath,amssymb,amsfonts}
\usepackage{graphicx}
\usepackage{textcomp}
\usepackage{xcolor}
\usepackage{booktabs}
 \usepackage{psfrag}
 \usepackage{caption}
\usepackage{subcaption}
\usepackage{balance}
\usepackage{multirow}
\usepackage{tabularx}
\usepackage{diagbox}

\usepackage[linesnumbered,ruled,vlined,algo2e]{algorithm2e}

\SetKwInput{KwInput}{Input}                
\SetKwInput{KwOutput}{Output}              

\usepackage{dsfont}
\usepackage{tikz}
\usetikzlibrary{shapes.geometric}
\usepackage{pgfplots}
\pgfplotsset{width=0.45\textwidth}

\usepackage{acronym}
\acrodef{OFO}{Online Feedback Optimization}

\usepackage{graphicx}
\graphicspath{{Figures/}}

\renewcommand{\d}{\mathrm{d}} 
\newcommand{\R}{\mathds{R}} 

\usepackage[noend]{algpseudocode}
\usepackage[nothing]{algorithm}

\algnewcommand\And{\textbf{and}}

\usepackage{url}
\makeatletter
\g@addto@macro{\UrlBreaks}{\UrlOrds}
\makeatother

\usepackage{array}
\newcolumntype{P}[1]{>{\centering\arraybackslash}p{#1}}

\begin{document}
\begin{frontmatter}

\title{Tuning of Online Feedback Optimization for setpoint tracking in centrifugal compressors\thanksref{footnoteinfo}}

\thanks[footnoteinfo]{Research supported by  Marie Curie Horizon Postdoctoral Fellowship project RELIC (grant no 101063948).}

\author[NTNU]{Marta Zagorowska} 
\author[OST]{Lukas Ortmann}
\author[ZHAW]{Alisa Rupenyan}
\author[ICL]{Mehmet Mercang{\"o}z} 
\author[NTNU]{Lars Imsland} 

\address[NTNU]{Department of Engineering Cybernetics, Norwegian University of Science and Technology, email: 
        {\tt \{marta.zagorowska,lars.imsland\}@ntnu.no}}
\address[OST]{Eastern Switzerland University of Applied Sciences, email: 
        {\tt lukas.ortmann@ost.ch}}
\address[ICL]{Department of Chemical Engineering, Imperial College London, email: 
        {\tt  m.mercangoz@imperial.ac.uk}}
\address[ZHAW]{ZHAW Centre for Artificial Intelligence, ZHAW Z\"{u}rich University of Applied Sciences, Switzerland, email: {\tt rupn@zhaw.ch}}

\begin{abstract}

Online Feedback Optimization (OFO) controllers steer a system to its optimal operating point by treating optimization algorithms as auxiliary dynamic systems. Implementation of OFO controllers requires setting the parameters of the optimization algorithm that allows reaching convergence, posing a challenge because the convergence of the optimization algorithm is often decoupled from the performance of the controlled system. OFO controllers are also typically designed to ensure steady-state tracking by fixing the sampling time to be longer than the time constants of the system. In this paper, we first quantify the impact of OFO parameters and the sampling time on the tracking error and number of oscillations of the controlled system, showing that adjusting them without waiting for steady state allows good tracking. We then propose a tuning method for the sampling time of the OFO controller together with the parameters to allow tracking fast trajectories while reducing oscillations. We validate the proposed tuning approach in a pressure controller in a centrifugal compressor, tracking trajectories faster than the time needed to reach the steady state by the compressor. The results of the validation confirm that simultaneous tuning of the sampling time and the parameters of OFO yields up to 87\% times better tracking performance than manual tuning based on steady state.

\end{abstract}

\begin{keyword}
Online Feedback Optimization, controller tuning, sampling time
\end{keyword}

\end{frontmatter}
\section{Introduction}
\label{sect:intro}
OFO controllers steer a system to a locally optimal operating point without explicitly solving a nonlinear constrained optimization problem \citep{Optimization_Hauswirth2021}, thus showing similarities with classic approaches like extremum seeking (ES) control \citep{Model_He2023}. Instead of introducing additional perturbations like ES, OFO ensures reaching the optimum by exploiting properties of feedback control and iterative optimization algorithms, typically based on gradients for fast convergence. Thus, OFO implementations require tuning of the parameters of the underlying optimization algorithm, such as the time step between iterations, the length of a step in a single iteration, as well as weighting matrices \citep{Optimization_Hauswirth2021}. Successful applications of OFO include electric grids \citep{Deployment_Ortmann2023} and compressor stations \citep{Online_Zagorowska202212} where the parameters of OFO were set experimentally. \cite{Data_Gil2023} proposed an iterative tuning method for the parameters of OFO in a distillation system, iteratively adjusting the step length and weights with a fixed sampling time until the desired convergence of OFO was reached. However, the choice of the sampling time and the proposed tuning method were based on expert knowledge on the effect of the controller on the distillation system and thus application specific. In this work, we propose a method for tuning simultaneously the time step and the parameters of OFO without using explicit knowledge about their impact on the responses of the system.  

Tuning of sampling time of controllers is classically done with respect to the timescales of the controlled system \citep{Automatic_Aastroem1984}. \cite{Data_Gil2023} chose the sampling time of OFO so that the controller runs on a slower timescale than the underlying dynamic system, thus ensuring timescale separation \citep{Timescale_Hauswirth2021}. \cite{Adaptive_Picallo2022} and \cite{Data_Gil2023} indicated that large sampling time may lead to suboptimal performance of the controller especially in OFO using gradient-based optimization algorithms. At the same time, \cite{Online_Belgioioso2022} have applied OFO with sampling time of minutes in optimization of building climate control operating on a timescale of hours, showing that OFO controllers can work without timescale separation. Thus, to find a trade-off between ensuring timescale separation and the performance of OFO based on gradient descent, we formulate a tuning optimization problem to find a sampling time that satisfies the requirements on the responses of the system. 

In this paper, we analyse the impact of parameters of OFO on the performance of the controlled system, indicating that adjustment of the sampling time allows shaping the response of the controlled system. We develop a tuning framework for OFO so that the response of the controlled system has the desired properties with respect to error tracking and oscillatory behaviour. The performance of tuning is shown in an OFO controller for tracking suction pressure in centrifugal compressors without timescale separation.

The paper is structured as follows. Section \ref{sec:OFODynamics} presents the optimization problem for OFO tuning, which is then applied in Section \ref{sec:CaseStudy} in a case study of compressor control. Section \ref{sec:ImpactAll} analyses the impact of parameters on OFO performance, while Section \ref{sec:Tuning} presents the tuning framework and validation results for the compressor. Section \ref{sect:concl} presents conclusions and directions for future work. 

\section{Online Feedback Optimization for dynamic systems}
\label{sec:OFODynamics}

\subsection{Dynamic system}
The controlled system is described by nonlinear dynamics:
\begin{equation}
\dot{x}(t)=f(x,u)
\label{eq:InitialDynamics}
\end{equation}
where $f:\R^{s}\times \R^{p}\rightarrow \R^{s}$ is continuously differentiable. The inputs $u\in\R^{p}$ are  constrained by the physics of the system, $-b_2\leq u \leq b_1$, $b_i\in\R_+^{p}$, $i=1,2$. The outputs $y\in\R^{n}$ are described by a continuously differentiable nonlinear mapping $y=g(x,u)$ and $g:\R^s\times\R^p\rightarrow\R^n$. We assume that for a constant $u$, the system \eqref{eq:InitialDynamics} reaches a steady state $x_s(u)$ such that $f(x_s,u)=0$. 
Then we have:
\begin{equation}
    y=g(x_s(u),u)=h(u)
    \label{eq:DynamicOutput}
\end{equation}
where $h:\R^s\rightarrow\R^n$ is a continuously differentiable nonlinear steady state mapping. We further assume that the outputs $y$ in \eqref{eq:DynamicOutput} are bounded for any bounded inputs $u$. We want to design an OFO controller so that the outputs \eqref{eq:DynamicOutput} track a setpoint $y_{sp}$.

\subsection{Online Feedback Optimization}
\label{sec:FO}
Online Feedback Optimization (OFO) is designed to solve problems of the form \citep{Adaptive_Picallo2022}:
\begin{subequations} \label{eqn:ProblemStatement}
\begin{align}
\min_{u\in\mathcal{U}, y\in\mathcal{Y}}& \quad \Phi(u,y)
    \label{eqn:CostFcn}\\
\text{subject to }    & y=h(u)\label{eq:mapping}
\end{align}
\end{subequations}
where $\Phi:\R^p\times\R^n\rightarrow\R$ is a continuously differentiable cost, and $
\mathcal{U}=\lbrace u\in\R^p:Au\leq b \rbrace $, $\mathcal{Y}=\lbrace y\in\R^n:Cy\leq d \rbrace$, where $A\in \R^{q\times p}$, $b\in\R^q$, $C\in\R^{l\times n}$, and $d\in\R^l$ are constant matrices \citep{Non_Haeberle2020}. \ac{OFO} iteratively updates $u$ in \eqref{eq:InitialDynamics} to make $y$ converge to a local optimum of \eqref{eqn:ProblemStatement}, even if $\Phi$ is non-convex \citep{Non_Haeberle2020}. 

\subsubsection{Optimization algorithm as a dynamic system}  
In this work, we use an OFO controller with a constant step size \(\alpha>0\) proposed by \cite{Non_Haeberle2020} and successfully deployed in a distribution grid by \cite{Deployment_Ortmann2023}: 
\begin{align}\label{eqn:Verena_feedback}
    u^{k+1} = u^k + \alpha\widehat{\sigma}_\alpha (u^k,y^k)
\end{align} 
where $y^k$ is the measured system output at time $k\Delta T$ with constant sampling time $\Delta T$ and \(\widehat{\sigma}_\alpha (u^k,y^k)\) is the minimizer of the constrained quadratic optimization problem:
\begin{subequations} \label{eqn:Verena_opt}
\begin{align}
    &\min_{w\in\mathbb{R}^p}\left|\left| w + G^{-1}H^\top(u^k)\nabla\Phi^\top(u^k,y^k)\right|\right|_{G}^2
    \label{eqn:Verena_sigma}\\
    &\text{subject to}\quad A\left(u^k+\alpha w\right)\leq b \label{eq:UProj}\\
    &\qquad\qquad\quad C \left(y^k+\alpha\nabla h(u^k) w\right)\leq d \label{eq:YProj}
\end{align}
\end{subequations}
where $w\in\mathbb{R}^p$ is an auxiliary decision variable of size $u$ and $G\in\mathbb{S}_+^p$ is a positive-definite matrix $p\times p$. The matrix \(H(u^k)^\top = \left[\mathbb{I}_p \ \nabla_u h(u^k)^\top\right]\), and $\nabla_{u} h(u^k)^\top$ is called \emph{input-output sensitivity}. We assume that the mapping $h$ is known, and $\nabla_u h$ can be computed analytically. More general approaches based on online gradient estimation were proposed by \cite{Model_He2023}. The matrix $\mathbb{I}_p$ is an identity matrix of size $p\times p$. The gradient of the objective function $\nabla\Phi(u,y)$ is:
\begin{equation}
    \nabla\Phi(u,y)=\begin{bmatrix}
    \frac{\partial \Phi}{\partial u_1} & \ldots & \frac{\partial \Phi}{\partial u_p} & \frac{\partial \Phi}{\partial y_1} & \ldots & \frac{\partial \Phi}{\partial y_n}
    \end{bmatrix}.
    \label{eq:GradientDef}
\end{equation}

\subsection{Online operation and tuning}

\subsubsection{Physical constraints}
For tracking, $\Phi(u,y):=\Phi(y,y_{sp})$, and if there are no output constraints, finding minimum in \eqref{eqn:Verena_sigma} allows steering the system to a local optimum. We assumed in Section \ref{sec:OFODynamics} that $y$ is bounded for any $u$, so we can remove \eqref{eq:YProj} from \eqref{eqn:Verena_opt}. We also note that $u$ is a control signal with physical limits, so we can remove \eqref{eq:UProj} from \eqref{eqn:Verena_sigma}, shifting the constraints from the optimization to the system. Then the control $u_{\text{applied}}^{k+1}$ is given by a saturation:
\begin{equation}
    u_{\text{applied}}^{k+1}=\max\lbrace -b_2,\min\lbrace b_1,u^{k+1} \rbrace \rbrace
    \label{eq:Saturation}
\end{equation}
where $u^{k+1}$ is obtained from \eqref{eqn:Verena_feedback} as:
\begin{equation}
    \hat{\sigma}_{\alpha}(u^k,y^k) =- G^{-1}H^\top(u^k)\nabla\Phi^\top(u^k,y^k).
    \label{eq:DefaultGradient}
\end{equation}
The physical constraints on the system transform \eqref{eqn:Verena_sigma} into OFO with the steepest descent algorithm in \eqref{eq:DefaultGradient} \citep{Optimization_Hauswirth2021}.

\subsubsection{Tuning}
The performance of OFO from \eqref{eqn:Verena_feedback} depends also on $\alpha$ and $G$ related to the optimization problem \eqref{eqn:Verena_sigma}. Assuming $G$ is a diagonal matrix, \cite{Data_Gil2023} propose an iterative tuning procedure, by increasing $\alpha$ and decreasing elements of $G$ with a fixed sampling time until the system starts to oscillate. Thanks to the transformation from \eqref{eq:DefaultGradient}, it is sufficient to tune the product $\nu:=\alpha G^{-1}$, reducing the number of tuned parameters to one. The value of $\nu$ affects how the control input changes from iteration $k$ to $k+1$ and corresponds to a step size in line search algorithms \citep{Optimization_Hauswirth2021}. Thus, a small $\nu$ will lead to slow convergence \citep{Nonlinear_Bertsekas2016} and a sluggish behaviour as the controller will need multiple iterations to reach the optimum. Conversely, a large $\nu$ may destabilize the system \citep{Non_Haeberle2020,Timescale_Hauswirth2021}. 
\section{Compressor control}
\label{sec:CaseStudy}

We design an OFO pressure controller in a compressor with nonlinear dynamics \citep{Experimental_Cortinovis2015}:
\begin{subequations} \label{eq:CompressorSystemCortinovisMultipleRecycle}
\begin{align}
\dot{p}_{s}&={}\frac{a_{01}^2}{V_{s}}(m_{\text{in}}-m)\\
\dot{p}_{d}&={}\frac{a_{01}^2}{V_{d}}(m-m_{\text{out}})\\
\dot{m} &={} \frac{A_{1}}{L_{c}}\left(\Pi\left(m,\omega\right) p_{s}-p_{d}\right)\label{eq:Pressures}\\
\dot{\omega} &={} \frac{1}{J}(\tau-\tau_{c})\label{eq:Torque}
\end{align}
\end{subequations}
where $p_{s}$ and $p_d$ are the suction and discharge suction pressures respectively, $a_{01}$, $V_{s}$, $A_{1}$, $L_{c}$, $J$ are constant parameters defining the geometry of the compressor, $m$ is the mass flow through the compressor, $\omega$ is the speed of the shaft of the compressor in rad\,s$^{-1}$, $\tau$ [Nm] is torque provided by a flow controller, $\tau_{c}$ is the reaction torque of the compressor, given as $\tau_c=\delta\omega m$ \citep{Drive_Gravdahl2002} with $\delta=0.00729$ capturing the internal geometry of the compressor \citep{Experimental_Cortinovis2015}. The compressor map $\Pi$ describes the pressure ratio across a compressor as a quadratic function of compressor mass flow and speed \citep{Real_Milosavljevic2020}. The value of $m_{\text{in}}$ and $m_{\text{out}}$ captures the external mass flows on the suction and discharge side, respectively. The mass flows depend on the pressures $p_s$ and $p_d$, and external pressures $p_\mathrm{in}$ and $p_\mathrm{out}$:
\begin{subequations} \label{eqn:mass-flows}
\begin{align}
    {m}_{\mathrm{in}}&= 0.4k_\mathrm{in}A_\mathrm{in}\sqrt{|p_\mathrm{in}-p_s|}\\
    {m}_{\mathrm{out}} &= 0.8k_\mathrm{out}A_\mathrm{out}\sqrt{|p_d-p_\mathrm{out}|}
\end{align}
\end{subequations}
where $A_\mathrm{in}$, $A_\mathrm{out}$, \(k_\mathrm{in}, k_\mathrm{out}\) are constant parameters~\citep{Real_Milosavljevic2020}. The initial condition for \eqref{eq:CompressorSystemCortinovisMultipleRecycle} is $p_s(0)=1.015$ bar, $p_d(0)=1.868$ bar, $m(0)=60.45$ kg s$^{-1}$, $\omega(0)=647.2$ rad s$^{-1}$ and corresponds to a steady-state torque $\tau(0)=323.6$ Nm. To obtain the steady state mapping from \eqref{eq:mapping}, we use $\omega=\frac{\tau}{\delta m}$ \citep{Drive_Gravdahl2002} in \eqref{eq:Pressures} to get $p_s=\frac{p_d}{\Pi\left(m,\frac{\tau}{\delta m}\right)}=h(\tau,m)$ where $m$ is measured and $p_d$ is obtained from \eqref{eqn:mass-flows}, taking $m_{\text{in}}=m_{\text{out}}$ for $p_{\text{in}}=1.05$ bar and $p_{\text{out}}=1.55$ bar. To put the compressor control problem in the framework from \eqref{eqn:ProblemStatement}, we set: $u=\tau$, $y=p_s$. The goal is to follow the desired suction pressure setpoint $p_{sd}$ with the objective function \eqref{eqn:CostFcn} formulated as:
\begin{equation}
\label{eq:CompressorObjective}
\Phi(p_s) = 0.01(p_s-p_{sd})^2.
\end{equation}
We assume no constraints on the suction pressure $p_s$, $\mathcal{Y}:=\R$. The torque $\tau$ is physically constrained by $b_1=1000$ Nm, $b_2=-300$ Nm in \eqref{eq:Saturation}. The bounds on the torque were chosen to ensure that the suction pressure is bounded for the given external conditions.

\section{Impact of OFO parameters}
\label{sec:ImpactAll}
\begin{figure*}
     \centering
         \psfrag{dt =0.5}[][]{\hspace{1.5mm}\scalebox{.45}{$\Delta T$\textsf{=0.5}} }
         \psfrag{dt =0.005}[][]{\hspace{1.5mm}\scalebox{.45}{$\Delta T$\textsf{=0.005}} }
         \psfrag{dt =5.0}[][]{\hspace{1.5mm}\scalebox{.45}{$\Delta T$\textsf{=5.0}} }
         \psfrag{dt =0.05}[][]{\hspace{1.5mm}\scalebox{.45}{$\Delta T$\textsf{=0.05}} }
         \psfrag{dt =50.0}[][]{\hspace{1.5mm}\scalebox{.45}{$\Delta T$\textsf{=50.0}} }
         \psfrag{v =1.0}[][]{\hspace{1.5mm}\scalebox{.45}{$\nu$\textsf{=1.0}} }
         \psfrag{v =10.0}[][]{\hspace{1.5mm}\scalebox{.45}{$\nu$\textsf{=10.0}} }
         \psfrag{v =1000.0}[][]{\hspace{1mm}\scalebox{.45}{$\nu$\textsf{=1000.0}} }
         \psfrag{v =0.1}[][]{\hspace{1.5mm}\scalebox{.45}{$\nu$\textsf{=0.1}} }
         \psfrag{v =0.001}[][]{\hspace{1.5mm}\scalebox{.45}{$\nu$\textsf{=0.001}} }

         \psfrag{vp}[][]{\tiny $\nu$}
         \psfrag{dt}[][]{\tiny $\Delta T$}
         \psfrag{p_s [bar]}[][]{\tiny $p_s$ \textsf{[bar]}}
         \psfrag{Torque [Nm]}[][]{\tiny\textsf{ Torque  [Nm]}}
        \psfrag{Time [s]}[][]{\tiny\textsf{Time [s]}}
    \psfrag{0}[][]{\scalebox{.4}{\textsf{0}}}
    \psfrag{200}[][]{\scalebox{.4}{\textsf{200}}}
    \psfrag{400}[][]{\scalebox{.4}{\textsf{400}}}
    \psfrag{800}[][]{\scalebox{.4}{\textsf{800}}}
    \psfrag{1000}[][]{\scalebox{.4}{\textsf{1000}}}
    \psfrag{50}[][]{\scalebox{.4}{\textsf{50}}}
    \psfrag{150}[][]{\scalebox{.4}{\textsf{150}}}
    \psfrag{250}[][]{\scalebox{.4}{\textsf{250}}}
    \psfrag{500}[][]{\scalebox{.4}{\textsf{500}}}
    \psfrag{750}[][]{\scalebox{.4}{\textsf{750}}}
    \psfrag{0.85}[][]{\scalebox{.4}{\textsf{0.85}}}
    \psfrag{0.95}[][]{\scalebox{.4}{\textsf{0.95}}}
    \psfrag{1.00}[][]{\scalebox{.4}{\textsf{1.00}}}
    \psfrag{0.90}[][]{\scalebox{.4}{\textsf{0.90}}}
    \psfrag{2}[][]{\scalebox{.4}{\textsf{2}}}
    \psfrag{4}[][]{\scalebox{.4}{\textsf{4}}}
    \psfrag{6}[][]{\scalebox{.4}{\textsf{6}}}
    \psfrag{8}[][]{\scalebox{.4}{\textsf{8}}}
    \psfrag{10}[][]{\scalebox{.4}{\textsf{10}}}
    \psfrag{12}[][]{\scalebox{.4}{\textsf{12}}}
    \psfrag{100}[][]{\scalebox{.4}{\textsf{100}}}
    \psfrag{25}[][]{\scalebox{.4}{\textsf{25}}}
    \psfrag{75}[][]{\scalebox{.4}{\textsf{75}}}
    \psfrag{125}[][]{\scalebox{.4}{\textsf{125}}}
    \psfrag{20}[][]{\scalebox{.4}{\textsf{20}}}
    \psfrag{30}[][]{\scalebox{.4}{\textsf{30}}}
    \psfrag{40}[][]{\scalebox{.4}{\textsf{40}}}
    \psfrag{60}[][]{\scalebox{.4}{\textsf{60}}}
    \psfrag{70}[][]{\scalebox{.4}{\textsf{70}}}
    \psfrag{80}[][]{\scalebox{.4}{\textsf{80}}}
    \psfrag{Setpoint}[][]{\scalebox{.5}{\textsf{Setpoint}}}

     \begin{subfigure}[b]{0.3\textwidth}
         \centering
         \includegraphics[width=\textwidth]{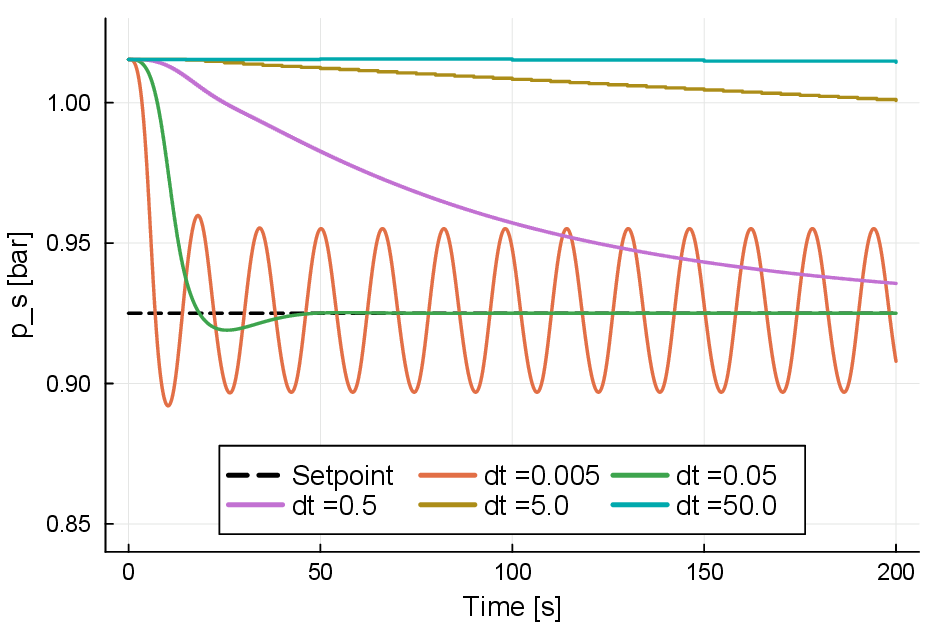}
         \caption{Impact of $\Delta T$ with $\nu=1$}
         \label{fig:Changetime}
     \end{subfigure}
     \hfill
     \begin{subfigure}[b]{0.3\textwidth}
         \centering
         \includegraphics[width=\textwidth]{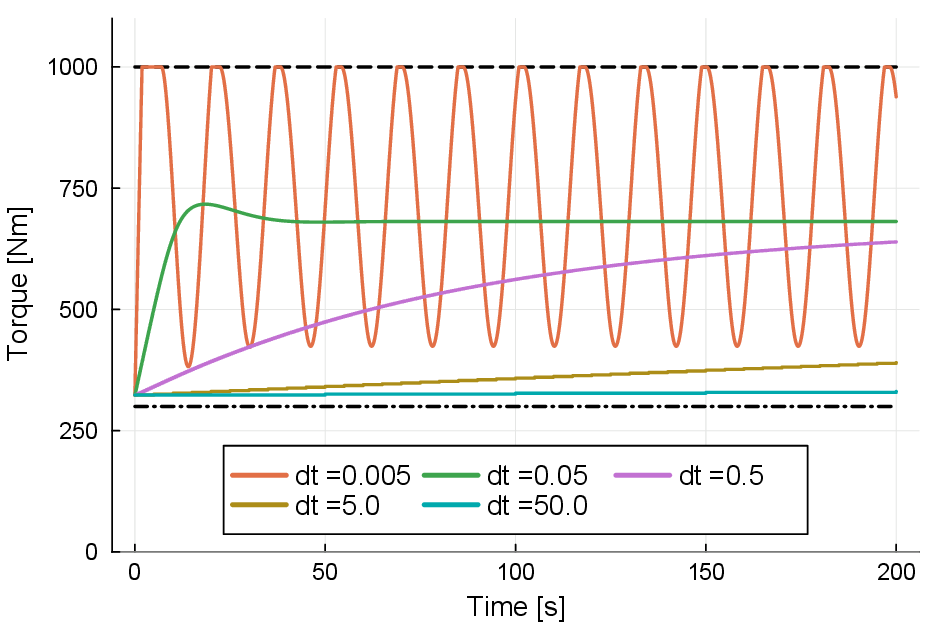}
         \caption{Impact of $\Delta T$ with $\nu=1$ on control}
         \label{fig:ControlTest_delta}
     \end{subfigure}
     \hfill
               \begin{subfigure}[b]{0.3\textwidth}
         \centering
         \includegraphics[width=\textwidth]{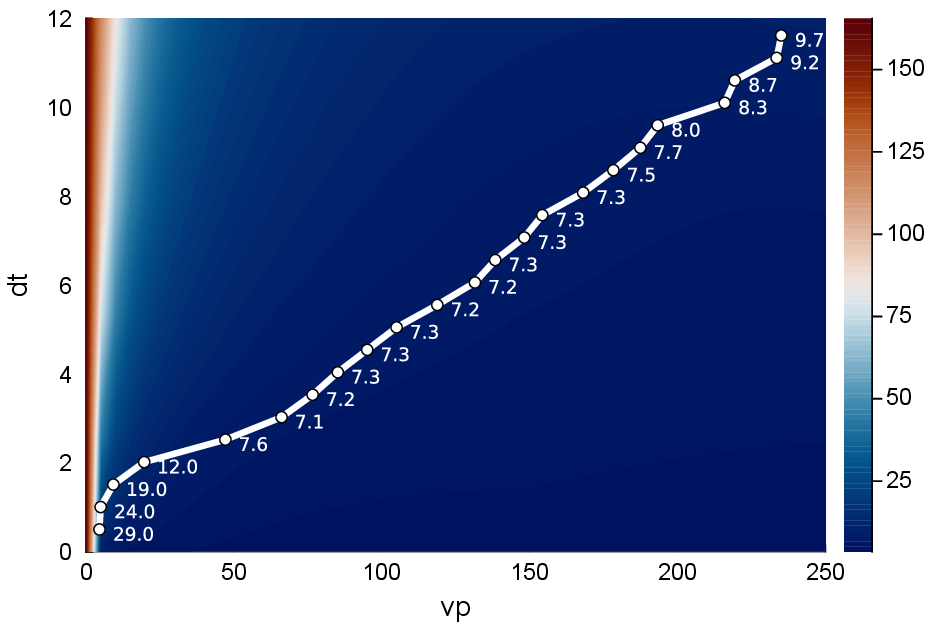}
         \caption{Error $\epsilon$}
         \label{fig:ContourError}
     \end{subfigure}
     \hfill
          \centering
               \begin{subfigure}[b]{0.3\textwidth}
         \centering
         \includegraphics[width=\textwidth]{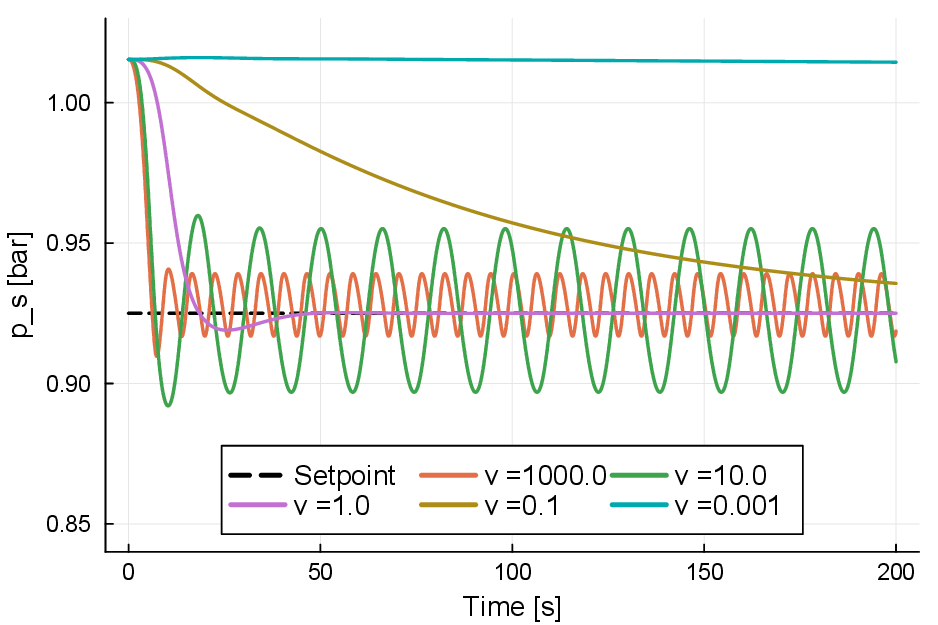}
         \caption{Impact of $\nu$ with $\Delta T=0.05$}
         \label{fig:Changealpha}
     \end{subfigure}
     \hfill
     \begin{subfigure}[b]{0.3\textwidth}
         \centering
         \includegraphics[width=\textwidth]{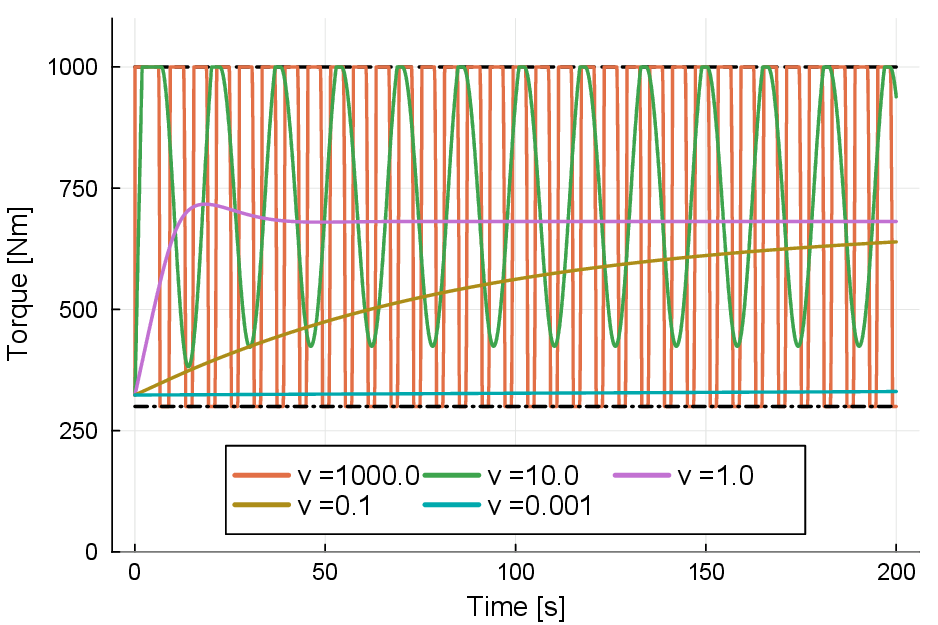}
         \caption{Impact of $\nu$ with $\Delta T=0.05$ on control}
         \label{fig:ControlTest_nu}
     \end{subfigure}
     \hfill
     \begin{subfigure}[b]{0.3\textwidth}
         \centering
         \includegraphics[width=\textwidth]{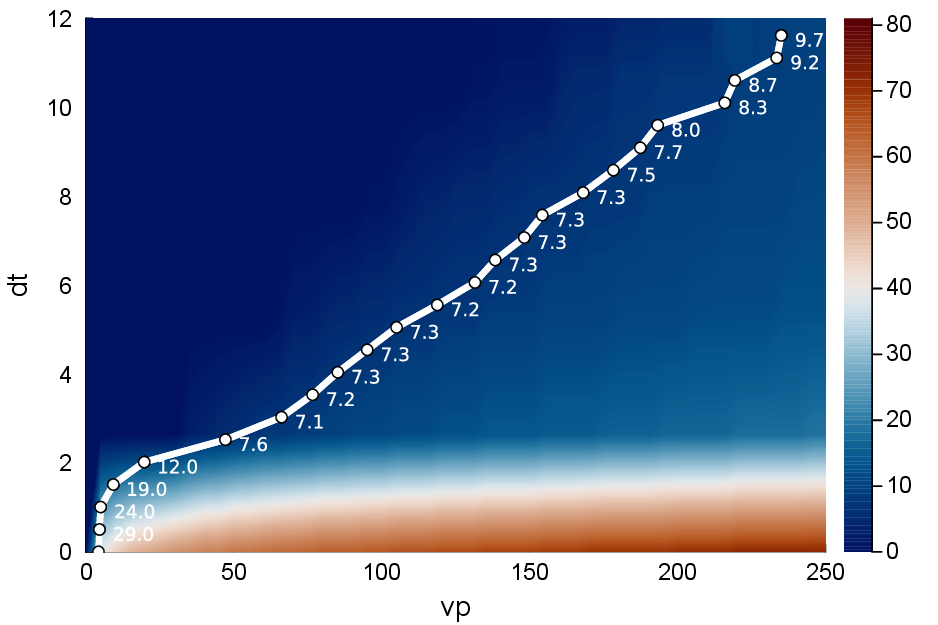}
         \caption{Oscillations $|F|$}
         \label{fig:ContourOscillations}
     \end{subfigure}
     \hfill
        \caption{Impact of parameters $\nu$ and $\Delta T$ (in s) on the performance of OFO with the objective \eqref{eq:CompressorObjective} and the corresponding control inputs, and a trade-off between the error and the number of oscillations as functions of parameters}
        \label{fig:ImpactParameters}
\end{figure*}

\label{sec:OFOObjectiveCstr}
We analyse the impact of $\nu$ and $\Delta T$ using \eqref{eq:GradientDef} on integrated squared error over a tuning horizon $t_F$:
\begin{equation}
    \epsilon(\nu,\Delta T):=\gamma_1\int\limits_0^{t_F} \left(p_s(\xi)-p_{sd}(\xi)\right)^2 \d \xi
    \label{eq:minError}
\end{equation}
 scaled by $\gamma_1=10^{-8}$, and on number of oscillations $|F(\nu,\Delta T)|$, based on counting zero crossings \citep{Detection_Thornhill2003}:
$F(\nu,\Delta T):=\lbrace t_k\in[0,t_F]: p_s(t_k)=p_{sd}(t_k)\rbrace$.
\subsection{Impact on performance}
\label{sec:Impact}

To show the relationship between $\nu$ and $\Delta T$, we analyse the contour plots of the error and the number of oscillations obtained for the constant pressure setpoint, with no setpoint for the torque as functions of the parameters (right column of Fig. \ref{fig:ImpactParameters}). The white line indicates where the error and the number of oscillations are equal. We also get that the smallest error is obtained for $\nu= 250$ and $\Delta T\leq 0.005$ (bottom right corner in Fig. \ref{fig:ContourError}) which corresponds to the largest number of oscillations (bottom right corner in Fig. \ref{fig:ContourOscillations}). Conversely, a small $\nu$ and a large $\Delta T$ give no oscillations (top left corner in Fig. \ref{fig:ContourOscillations}), but increase the error (top left corner in Fig. \ref{fig:ContourError}). The values annotating the intersection line indicate that similar performance can be obtained for $\nu\in[75,175]$ and $\Delta T\in[3,9]$ s, with both error and oscillations approximately 7.2, which suggests that the parameters should be tuned simultaneously.

\subsection{Impact on system}

Figure \ref{fig:ImpactParameters} shows the impact of parameters $\Delta T$ (Fig. \ref{fig:Changetime}) and $\nu$ (Fig. \ref{fig:Changealpha}) on the performance of the OFO controller for a constant setpoint. The oscillatory trajectories in Fig. \ref{fig:ImpactParameters} show the numerical connection between $\nu$ and the derivatives in \eqref{eq:DefaultGradient}. From \eqref{eq:DefaultGradient}, we see that:
\begin{equation}
    u^{k+1}-u^k = - \nu H^\top(u^k)\nabla\Phi^\top(u^k,y^k).
    \label{eq:ControlIncrement}
\end{equation}  
The parameter $\nu=1$ in Fig. \ref{fig:Changetime}, so the controller reaches steady state when the optimum is reached and $H^\top(u^k)\nabla\Phi^\top(u^k,y^k)$ becomes close to zero (vanishing oscillations for $\Delta T\leq 0.05$ in Fig. \ref{fig:Changetime}). If $\nu<1$, the derivatives are mitigated, $\nu\|H^{\top}\nabla\Phi^{\top}\|\leq \|H^{\top}\nabla\Phi^{\top}\|$, and the controller increments in \eqref{eq:ControlIncrement} are small, leading to sluggish behaviour ($\nu\leq 0.1$ in Fig. \ref{fig:Changealpha}). Conversely,  $\nu>1$ intensifies the effect of the derivatives, $\nu\|H^{\top}\nabla\Phi^{\top}\|\geq \|H^{\top}\nabla\Phi^{\top}\|$, leading to oscillatory behaviour ($\nu\geq 10$ in Fig. \ref{fig:Changealpha}), in line with the interpretation as a step size in line search algorithms \citep[p.31]{Nonlinear_Bertsekas2016}. 

\section{Tuning and validation}
\label{sec:Tuning}
\subsection{Optimization problem}
To tune OFO, we solve the optimization problem:
\begin{subequations} \label{eq:MaxTimeCstrError}
\begin{align}
 \underset{\nu,\Delta T}{\text{max}}& {}\quad\Delta T\label{eq:MaxTimeObj}&\\
\text{subject to:}    &\quad  \epsilon(\nu,\Delta T) \leq \beta_1&{}\label{eq:minErrorCstr}\\
&\quad   |F(\nu,\Delta T)|\leq \beta_2.&\label{eq:minOscCstr}
\end{align}
\end{subequations}
The objective \eqref{eq:MaxTimeObj} promotes a large sampling time so that the underlying dynamics reaches the steady state within $\Delta T$. The constraint \eqref{eq:minErrorCstr} reinforces the reference tracking in OFO from \eqref{eq:CompressorObjective} by restricting the time horizon to $t_F$. The constraint \eqref{eq:minOscCstr} enforces desired properties of the controller without affecting reference tracking. Both constraints are adjusted by choosing $\beta=[\beta_1,\beta_2]$, as suggested by Fig. \ref{fig:ImpactParameters} (right-most column), with $\beta_1$ as a threshold for the error and $\beta_2$ for the number of oscillations. To solve \eqref{eq:MaxTimeCstrError}, we use a constant setpoint $p_{sd}^{\text{const}}=0.925$ bar and two trajectories: a truncated sinusoidal signal:
\begin{equation}
    p^{\text{sine}}_{sd}(t)=\max\lbrace 0.94,0.95+0.05\sin(0.04t) \rbrace
    \label{eq:SineSP}
\end{equation}
and a step signal
\begin{equation}
    p^{\text{step}}_{sd}(t)=\begin{cases} 0.93\text{ bar if }t\in[75,125] \text{ s }\\
    0.98 \text{ bar if }t\in[0,75)\cup(125,150]  \text{ s. }\\
    0.95 \text{ bar if }t\in(150,t_F]  \text{ s }
    \end{cases}
    \label{eq:StepSP}
\end{equation}
The tuned values from solving \eqref{eq:MaxTimeCstrError} were compared with a default OFO controller, with $\Delta T=47.5$ s  ensuring timescale separation and corresponding to the settling time of the compressor from \eqref{eq:CompressorSystemCortinovisMultipleRecycle} (within 5\% of the final value \citep{controlsystems_jl}) linearized around the operating points for $\tau=323.6$ Nm.

The system \eqref{eq:CompressorSystemCortinovisMultipleRecycle} with the controller \eqref{eq:Saturation} was simulated using \texttt{OrdinaryDiffEq.jl} \citep{DifferentialEquations.jl-2017}. The derivatives $\nabla h$ and $\nabla \Phi$ for solving \eqref{eqn:Verena_sigma} were obtained using \texttt{Zygote v0.6.66} \citep{Zygote.jl-2018}. The optimization problem was implemented in an open source programming language, Julia 1.9.0, with Windows (x86\_64-w64-mingw32) CPU: 8 × 11th Gen Intel(R) Core(TM) i7-1165G7 @ 2.80GHz. The unconstrained optimization problem \eqref{eqn:Verena_sigma} was solved using \texttt{OSQP v0.8.0} \citep{osqp_Stellato2020}. To overcome potential non-differentiability \citep{Continuity_Bachtiar2016} when solving \eqref{eq:MaxTimeCstrError} with respect to the sampling time, we use a derivative-free solver, with a Julia interface \texttt{NOMAD.jl} \citep{montoison-pascal-salomon-nomad-2020} to NOMAD 4.3.1 \citep{nomad4paper}. The parameter $\nu\in [0,10^3]$ is bounded for scaling purposes, and $\Delta T\in[5\times 10^{-3},t_F/2]$ reflects the physical setup of the compressor defining the smallest and the largest sampling time available.  An analysis of the impact of number of iterations $t_f/\Delta T$ on the performance of OFO was done by \cite{Model_He2023}. We chose $t_F=200$ s to reflect typical operating times of centrifugal compressors \citep{Experimental_Cortinovis2015}.

{\scriptsize
\begin{table*}[!tbp]
\centering
\caption{Tuned values for the three setpoints, with parameters in \textbf{bold} chosen for validation. The error $\epsilon$ is reported as absolute value and improvement from to the manual tuning}
\label{tbl:TunedValues}
\resizebox{\textwidth}{!}{%
\begin{tabular}{l|llll|llll|llll|l}
              & \multicolumn{4}{l|}{Constant setpoint}                     & \multicolumn{4}{l|}{Step setpoint}                           & \multicolumn{4}{l}{Sinusoidal setpoint}         &              \\
$\beta$       & $\nu$          & $\Delta T$ [s] & $\epsilon$     & $|F|$       & $\nu$          & $\Delta T$ [s] & $\epsilon$        & $|F|$      & $\nu$          & $\Delta T$ [s] & $\epsilon$          & $|F|$     & \\ \hline
Initial & 0.1            & 50             & 163.4 (-297\%)     & 0           & 0.1            & 50             & 85.47 (-85\%)        & 0          & 0.1            & 50             & 60.06 (-99\%)         & 0      &    \\
150, 50       & 40             & 99             & 136.03 (-231\%)    & 0           & 0              & 99             & 86.29 (-87\%)        & 0          & 0              & 99             & 60.75 (-101\%)         & 0     &     \\
37.5, 50      & 196.1          & 45.89          & 37.5 (9\%)      & 0           & 183.1          & 74.95          & 37.3  (19\%)        & 2          & 85.1           & 99             & 37.26 (-23\%)         & 0       &   \\
18.75, 25      & 202.1          & 22.88          & 18.7 (55\%)      & 1           & 200.1          & 22.79          & 18.75 (59\%)        & 4          & \textbf{199.1} & \textbf{80}    & \textbf{18.69} (38\%) & \textbf{0}& S3 \\
9, 12         & 223.1          & 10.9           & 8.96  (78\%)     & 9           & \textbf{207.1} & \textbf{12.88} & \textbf{8.97} (81\%) & \textbf{9} & 229.1          & 10.37          & 9 (70\%)             & 5        &  S2 \\
6, 20         & \textbf{468.1} & \textbf{5.78}  & \textbf{6} (85\%) & \textbf{18} & 237.1          & 7.63           & 5.96 (87\%)         & 12         & 278.1          & 7.52           & 6  (80\%)            & 9      &  S1  \\
Manual & \textbf{150}            & \textbf{47.5}           & \textbf{41.13} (-)     & \textbf{0}           & 300            & 47.5           & 46.76 (-)        & 2          & 175            & 47.5           & 30.21 (-)         & 2     &   SM 
\end{tabular}%
}
\end{table*}
}

\begin{figure*}[!tbp]
     \centering
         \psfrag{p_s [bar]}[][]{\tiny $p_s$ \textsf{[bar]}}
        \psfrag{Torque [Nm]}[][]{\tiny\textsf{ Torque  [Nm]}}
        \psfrag{Time [s]}[][]{\tiny\textsf{Time [s]}}
            \psfrag{p_s [bar]}[][]{\tiny $p_s$ \textsf{[bar]}}
     \psfrag{b =[18.75, 25.0]}[][]{\scalebox{.45}{$\beta $\textsf{=[18.75, 25.0]}} }
     \psfrag{b =[150.0, 50.0]}[][]{\scalebox{.45}{$\beta$\textsf{=[150.0, 50.0]}} }
     \psfrag{b =[9.0, 12.0]}[][]{\scalebox{.45}{$\beta$\textsf{=[9.0, 12.0]}} }
     \psfrag{b =[37.5, 50.0]}[][]{\scalebox{.45}{$\beta $\textsf{=[37.5, 50.0]}} }
     \psfrag{b =[6.0, 20.0]}[][]{\scalebox{.45}{$\beta $\textsf{=[6.0, 20.0]}} }
    \psfrag{Setpoint}[][]{\scalebox{.5}{\textsf{Setpoint}}}
    \psfrag{Initial guess}[][]{\scalebox{.5}{\textsf{Initial guess}}}
    \psfrag{Manual}[][]{\scalebox{.5}{\textsf{Manual}}}

    \psfrag{0}[][]{\scalebox{.4}{\textsf{0}}}
    \psfrag{200}[][]{\scalebox{.4}{\textsf{200}}}
    \psfrag{400}[][]{\scalebox{.4}{\textsf{400}}}
    \psfrag{800}[][]{\scalebox{.4}{\textsf{800}}}
    \psfrag{1000}[][]{\scalebox{.4}{\textsf{1000}}}
    \psfrag{50}[][]{\scalebox{.4}{\textsf{50}}}
    \psfrag{150}[][]{\scalebox{.4}{\textsf{150}}}
    \psfrag{250}[][]{\scalebox{.4}{\textsf{250}}}
    \psfrag{500}[][]{\scalebox{.4}{\textsf{500}}}
    \psfrag{750}[][]{\scalebox{.4}{\textsf{750}}}
    \psfrag{0.85}[][]{\scalebox{.4}{\textsf{0.85}}}
    \psfrag{0.95}[][]{\scalebox{.4}{\textsf{0.95}}}
    \psfrag{1.00}[][]{\scalebox{.4}{\textsf{1.00}}}
    \psfrag{0.90}[][]{\scalebox{.4}{\textsf{0.90}}}
    \psfrag{100}[][]{\scalebox{.4}{\textsf{100}}}

     \begin{subfigure}[b]{0.3\textwidth}
    \centering
    \includegraphics[width=\textwidth]{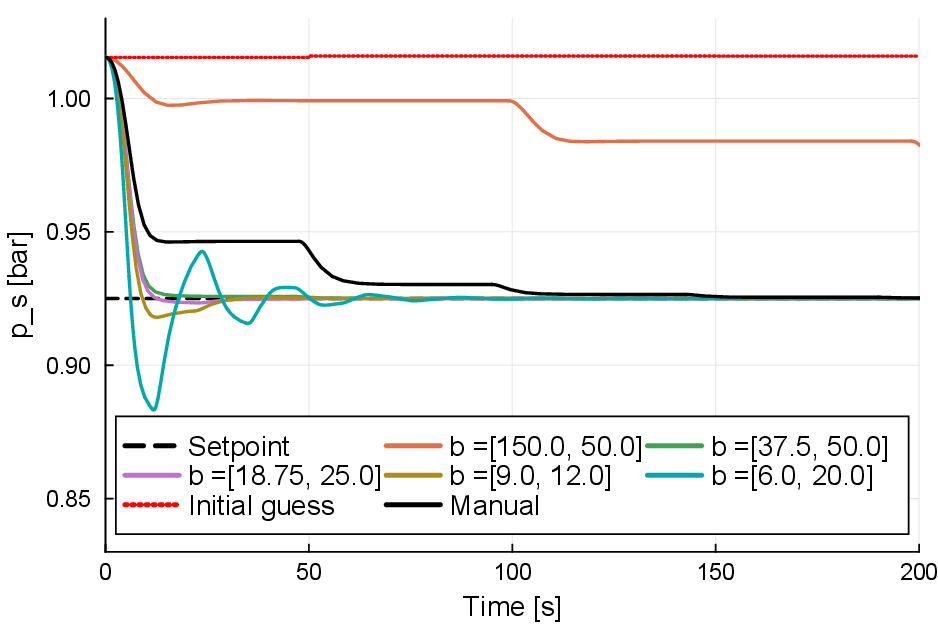}
    \caption{Constant setpoint}
    \label{fig:ResultTuningConstPressure}
     \end{subfigure}
     \hfill
     \begin{subfigure}[b]{0.3\textwidth}
         \centering
         \includegraphics[width=\textwidth]{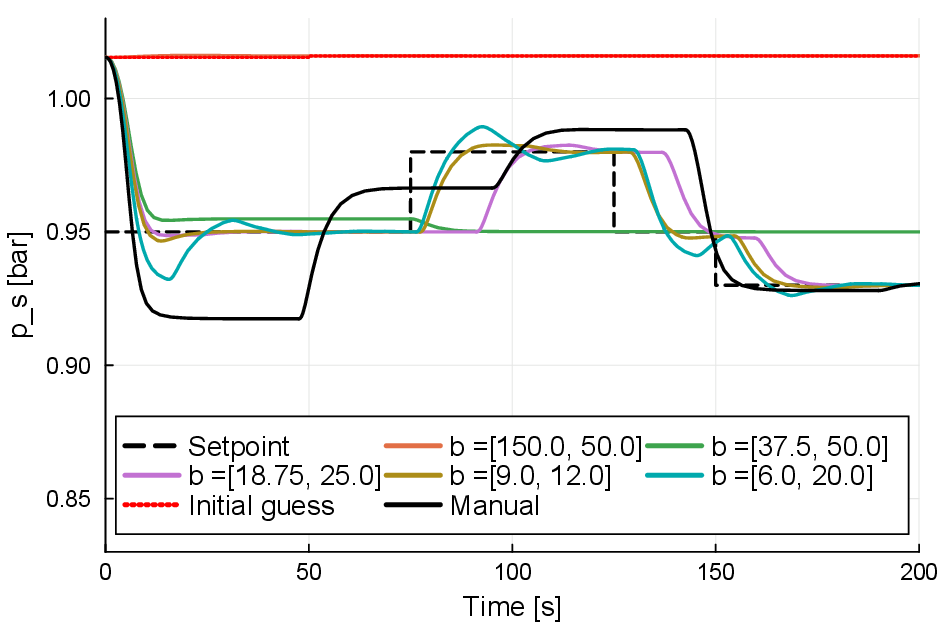}
         \caption{Step setpoint}
         \label{fig:ResultTuningStepPressure}
     \end{subfigure}
     \hfill
     \begin{subfigure}[b]{0.3\textwidth}
         \centering
         \includegraphics[width=\textwidth]{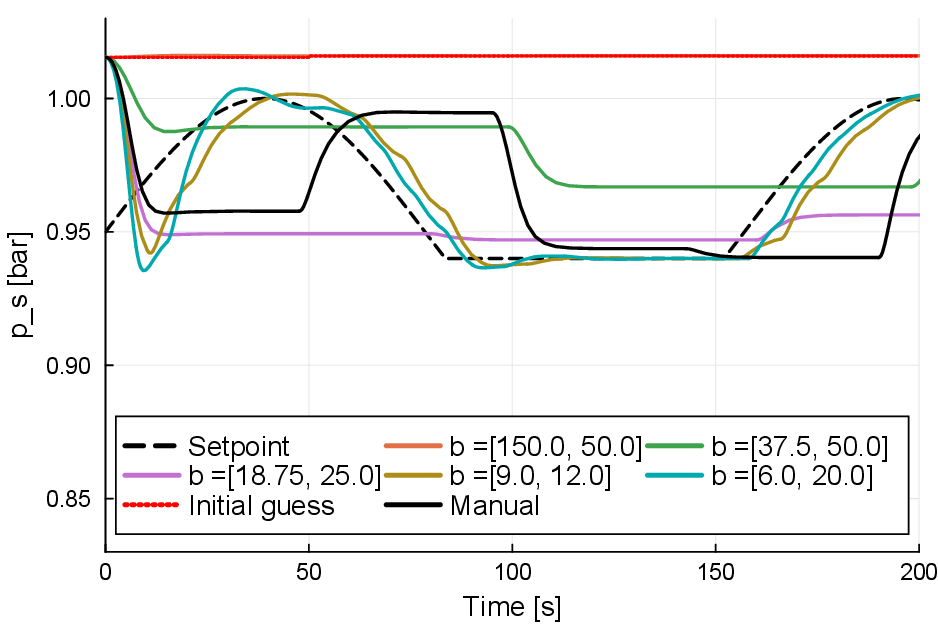}
         \caption{Sinusoidal setpoint}
         \label{fig:ResultTuningSinePressure}
     \end{subfigure}
     \hfill
     \centering
     \begin{subfigure}[b]{0.3\textwidth}
    \centering
    \includegraphics[width=\textwidth]{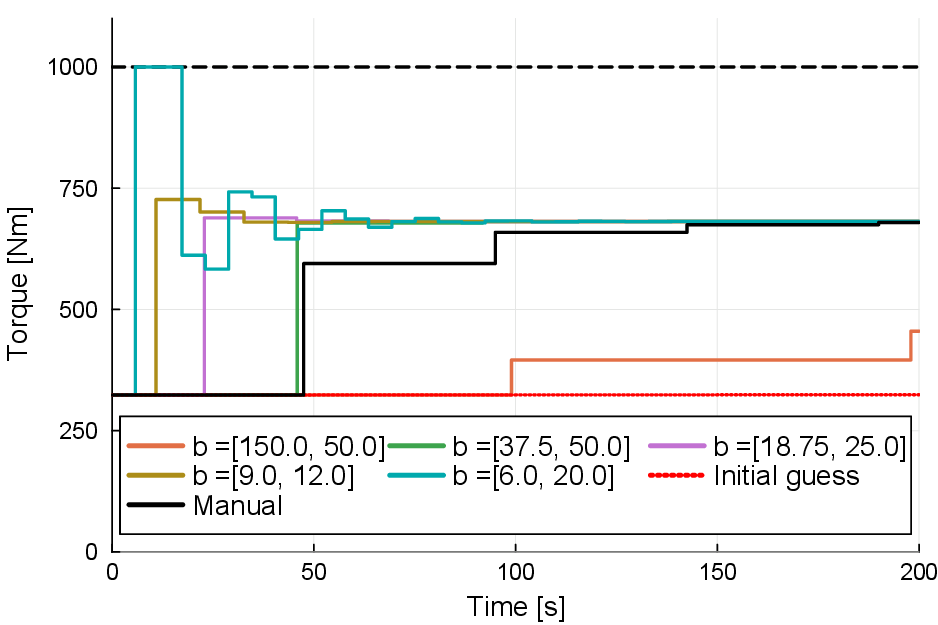}
    \caption{Control inputs for constant setpoint}
    \label{fig:ResultTuningConstControl}
     \end{subfigure}
     \hfill
     \begin{subfigure}[b]{0.3\textwidth}
         \centering
         \includegraphics[width=\textwidth]{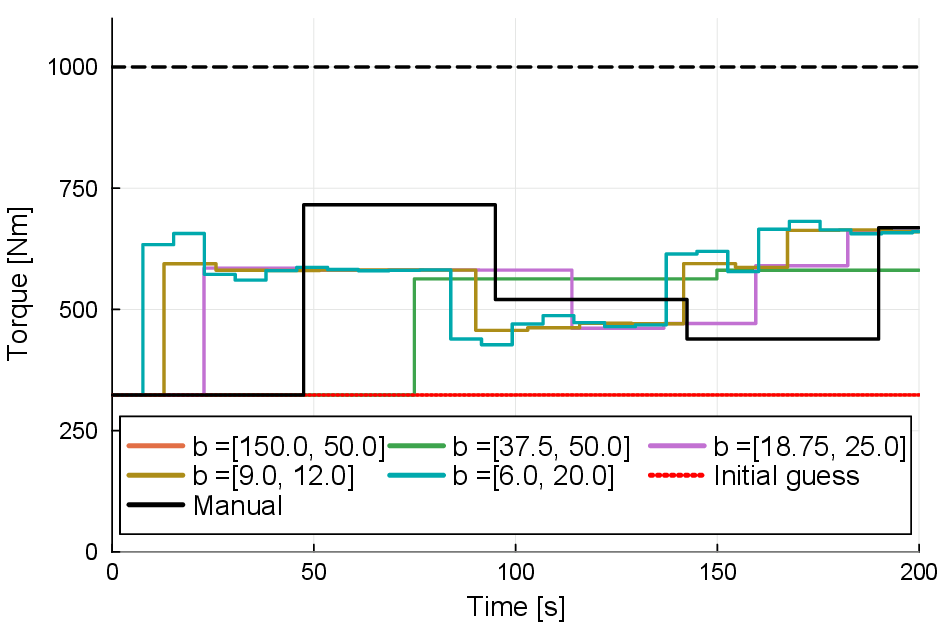}
         \caption{Control inputs for step setpoint}
         \label{fig:ResultTuningStepControl}
     \end{subfigure}
     \hfill
     \begin{subfigure}[b]{0.3\textwidth}
         \centering
         \includegraphics[width=\textwidth]{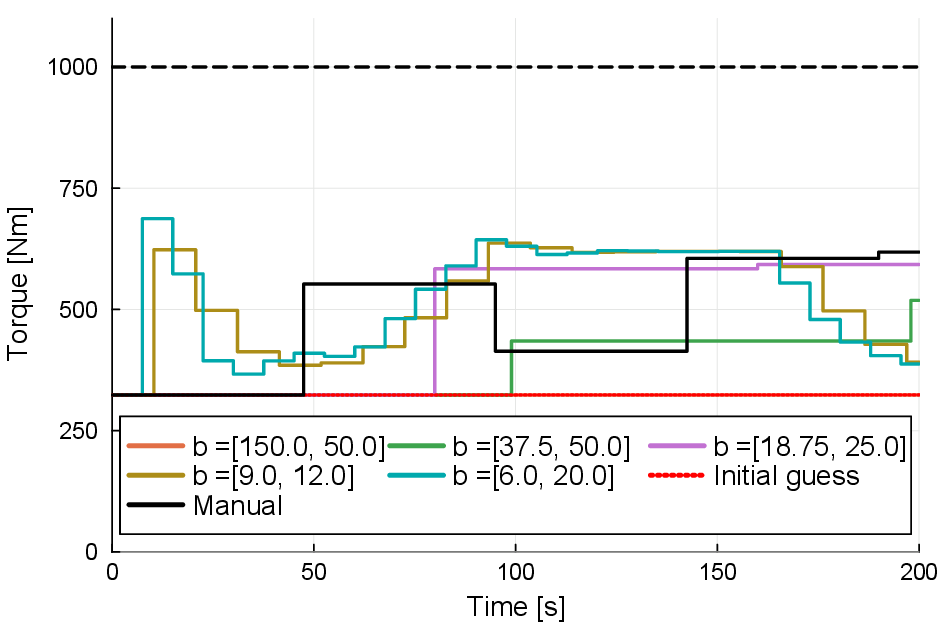}
         \caption{Control inputs for sinusoidal setpoint}
         \label{fig:ResultTuningSineControl}
     \end{subfigure}
     \hfill
        \caption{Results of tuning for different setpoints}
        \label{fig:ResultsSetpoints}
\end{figure*}

\subsection{Results for tuning}

The results of tuning OFO for the compressor \eqref{eq:CompressorSystemCortinovisMultipleRecycle} are shown in Fig. \ref{fig:ResultsSetpoints} with the initial guess for parameters $\Delta T=50$, $\nu=0.1$ in red, the setpoint in dashed black (top row), and the bounds for the torque in black (bottom row). The initial guess for $\nu$ and $\Delta T$ was chosen following the recommendation from \cite{Data_Gil2023}, as indicated in Table \ref{tbl:TunedValues}. The values corresponding to the default OFO controller are in solid black lines. To achieve the desired performance, the values of $\beta$ in \eqref{eq:MaxTimeCstrError} were applied in decreasing order, starting from $\beta=[150,50]$, as suggested by Fig. \ref{fig:ImpactParameters}. Figure \ref{fig:ResultsSetpoints} shows that for $\beta_1=150$ and $\beta_2=50$, the constraints on the error and the number of oscillations have little impact on shaping the response ($\Delta T=99$ in Table \ref{tbl:TunedValues}), and lead to a sluggish controller ($\beta_2=150$ in Fig. \ref{fig:ResultTuningConstControl}). Putting higher priority on the oscillations with $\beta_2\leq 25$, and decreasing $\beta_1$ from 37.5 to nine allows achieving better tracking with damped oscillations, at the expense of decreased sampling time ($\Delta T\leq 12$ in Table \ref{tbl:TunedValues}). The oscillatory behaviour of the control signal in Fig. \ref{fig:ResultTuningConstControl} confirms the importance of preserving the bounds on control signal in \eqref{eq:Saturation}. Restricting the number of oscillations, $\beta_2=12$, yield a less aggressive response (Fig. \ref{fig:ResultTuningConstControl}), at the expense of increased error.

\label{sec:Validation}
\begin{figure*}[!tbp]
     \centering
         \psfrag{p_s [bar]}[][]{\tiny $p_s$ \textsf{[bar]}}
        \psfrag{Torque [Nm]}[][]{\tiny\textsf{ Torque  [Nm]}}
        \psfrag{Time [s]}[][]{\tiny\textsf{Time [s]}}
            \psfrag{p_s [bar]}[][]{\tiny $p_s$ \textsf{[bar]}}
    \psfrag{Setpoint}[][]{\scalebox{.5}{\textsf{Setpoint}}}
    \psfrag{Set 1}[][]{\scalebox{.5}{\textsf{Set 1}}}
        \psfrag{Set 2}[][]{\scalebox{.5}{\textsf{Set 2}}}
    \psfrag{Set 3}[][]{\scalebox{.5}{\textsf{Set 3}}}

    \psfrag{Manual}[][]{\scalebox{.5}{\textsf{Manual}}}
    \psfrag{600}[][]{\scalebox{.4}{\textsf{600}}}
    \psfrag{0}[][]{\scalebox{.4}{\textsf{0}}}
    \psfrag{200}[][]{\scalebox{.4}{\textsf{200}}}
    \psfrag{400}[][]{\scalebox{.4}{\textsf{400}}}
    \psfrag{800}[][]{\scalebox{.4}{\textsf{800}}}
    \psfrag{1000}[][]{\scalebox{.4}{\textsf{1000}}}
    \psfrag{0.90}[][]{\scalebox{.4}{\textsf{0.90}}}
    \psfrag{0.93}[][]{\scalebox{.4}{\textsf{0.93}}}
    \psfrag{0.96}[][]{\scalebox{.4}{\textsf{0.96}}}
    \psfrag{0.99}[][]{\scalebox{.4}{\textsf{0.99}}}
    \psfrag{1.02}[][]{\scalebox{.4}{\textsf{1.02}}}
    \psfrag{250}[][]{\scalebox{.4}{\textsf{250}}}
    \psfrag{500}[][]{\scalebox{.4}{\textsf{500}}}
    \psfrag{750}[][]{\scalebox{.4}{\textsf{750}}}
    \psfrag{0.85}[][]{\scalebox{.4}{\textsf{0.85}}}
    \psfrag{0.95}[][]{\scalebox{.4}{\textsf{0.95}}}
    \psfrag{1.00}[][]{\scalebox{.4}{\textsf{1.00}}}

     \begin{subfigure}[b]{0.3\textwidth}
    \centering
    \psfrag{p_s [bar]}[][]{\tiny $p_s$ \textsf{[bar]}}
    \includegraphics[width=\textwidth]{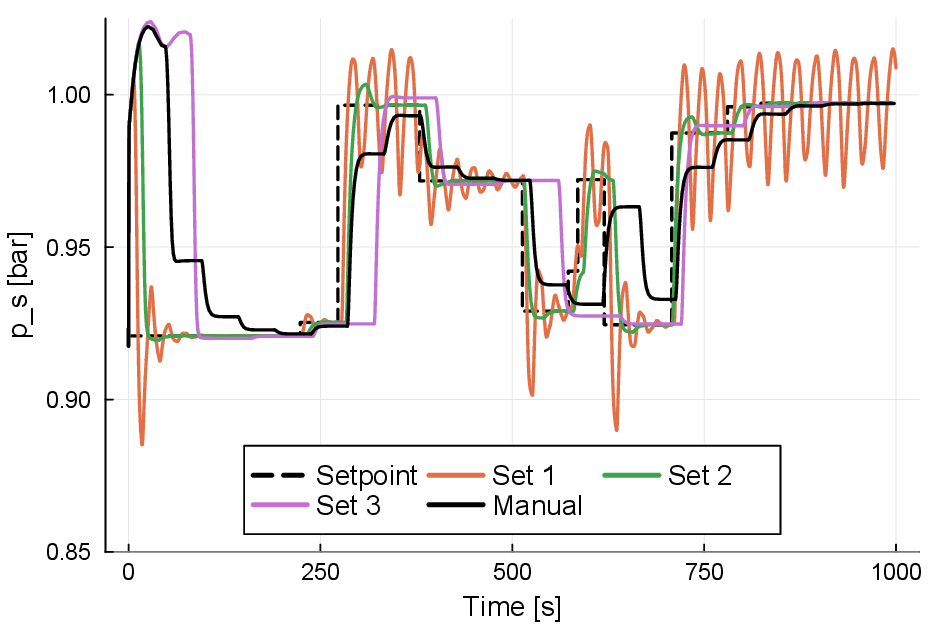}
    \caption{Validation for a step trajectory}
    \label{fig:ValidationStepResponse}
     \end{subfigure}
     \hfill
     \begin{subfigure}[b]{0.3\textwidth}
         \centering
         \includegraphics[width=\textwidth]{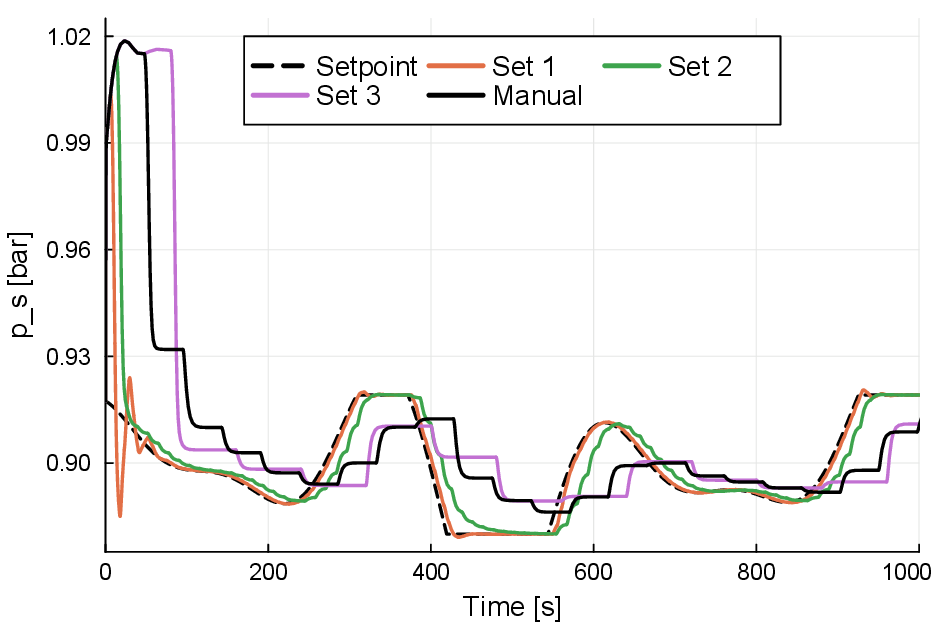}
         \caption{Validation for a sinusoidal trajectory}
         \label{fig:ValidationSine}
     \end{subfigure}
     \hfill
          \begin{subfigure}[b]{0.3\textwidth}
    \centering
    \psfrag{p_s [bar]}[][]{\tiny $p_s$ \textsf{[bar]}}
    \includegraphics[width=\textwidth]{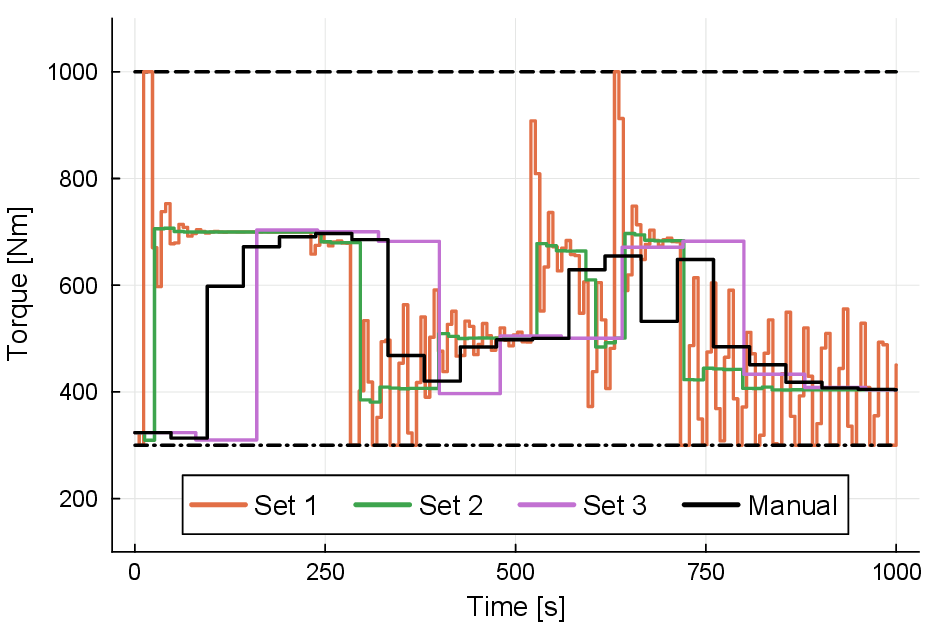}
    \caption{Control inputs for step validation}
    \label{fig:ValidationStepControl}
     \end{subfigure}
        \caption{Validation of the tuning parameters corresponding to constant tuning trajectory (Set 1), step tuning trajectory (Set 2), and sinusoidal tuning trajectory (Set 3) for two different setpoint trajectories}
        \label{fig:Validation}
\end{figure*}

\subsection{Results for validation}

To validate the tuning results, we track two trajectories for $p_{sd}$: a sinusoidal trajectory and a step trajectory (dashed black in Fig. \ref{fig:Validation}) reflecting possible setpoints for centrifugal compressors.  The three sets of parameters for validation, Set 1, Set 2, and Set 3 from Table \ref{tbl:TunedValues}, correspond to the smallest error obtained for constant setpoint, the smallest number of oscillations with the error below 10 for the step setpoint, and the largest $\Delta T$ and decreased error for sinusoidal tuning trajectory, respectively. The manual tuning for constant setpoint was also chosen because it gave the largest decrease in the error and zero oscillations compared to manual tuning with the step or sinusoidal setpoints. The initial condition for validation was set to $p_s(0)=0.91745$ bar, $p_d(0)=2$ bar, $m(0)=80$ kg s$^{-1}$, and $\omega(0)=700.5$ rad s$^{-1}$. 

The results of validation are shown in Fig. \ref{fig:Validation} and Table \ref{tbl:ErrorValidation} and confirm that solving \eqref{eq:MaxTimeCstrError} to find optimal parameters allows shaping the response of the system. Comparing the impact of the timestep $\Delta T$ on the performance, we see that a large value allows avoiding oscillations (Set 3 and Manual in Fig. \ref{fig:ValidationStepResponse}). Thus, the parameters obtained from \eqref{eq:MaxTimeCstrError} for the sinusoidal setpoint (Set 3) are comparable with the manual tuning preserving timescale separation. At the same time, a large $\Delta T$ leads to a sluggish controller and introduces delays (Fig. \ref{fig:ValidationStepControl}). The performance for the parameters in Set 1 and Set 2 is further defined by the value of $\nu$. The oscillatory behaviour is due to including the setpoint in the calculations of the gradients in \eqref{sec:OFOObjectiveCstr}. From \eqref{eq:CompressorObjective}, we have $\partial^2 \Phi/ \partial y\partial y_{sp}=-0.02y_{sp}$ and thus a large change in the setpoint $y_{sp}$ corresponds to a large change in the gradient $\partial \Phi/ \partial y$ in \eqref{eq:DefaultGradient}. In Set 1, combining the large change in the gradient due to the setpoint with a large $\nu$ in \eqref{eqn:Verena_feedback} leads to the controller changing significantly within a small $\Delta T$ and introduces oscillations. At the same time, the parameters from Set 1 allow following the sinusoidal trajectory with the smallest error (first row in Table \ref{tbl:ErrorValidation}) because there are no abrupt changes in the setpoint trajectory.

\subsection{Discussion and recommendations}
The results confirm a nonlinear relationship between the sampling time $\Delta T$ and the step size $\nu$ in \eqref{eqn:Verena_opt}, suggested by \cite{Online_Belgioioso2022}. Instead of iterative tuning by adjusting directly $\Delta T$ and $\nu$, which do not have explicit interpretation in terms of the responses of the system, solving the optimization problem \eqref{eq:MaxTimeCstrError} enables adjusting the thresholds $\beta_1$ and $\beta_2$ directly related to the error and the number of oscillations. As a possible choice of $\beta_1$, we can take the error \eqref{eq:minError} obtained for the initial steady state and a chosen tuning trajectory, $
    \beta_1=\gamma_1\int\limits_0^{t_F} (p_s(0)-p_{sd}(\xi))^2 \d \xi.$
The initial value for $\beta_2$ can be chosen with respect to the tuning horizon $t_F$, for instance $t_F/2$, allowing one zero crossing per two time units. Then both $\beta_1$ and $\beta_2$ can be decreased until the desired performance is reached. 

{\scriptsize
\begin{table}[!tbp]
\centering
\caption{Error \eqref{eq:minError} for the validation trajectories, as absolute values and percentage improvement w.r.t. manual tuning (SM)}
\label{tbl:ErrorValidation}
\begin{tabular}{@{}l|ll@{}}
\diagbox[width=8.5em]{Tuning}{Validation} & Step  & Sinusoidal \\ \midrule
Set 1 (S1)  & 29.98 (75\%) & 5.63 (94\%)\\
Set 2 (S2)      & 41.51 (65\%) & 17.94 (80\%) \\
Set 3 (S3)      & 201.07 (-69\%)& 139.51 (-53\%)\\
Manual (SM)      & 119.2 (-)& 91.27 (-)
\end{tabular}%
\end{table}
}

\section{Conclusions and future works}
\label{sect:concl}

Online Feedback Optimization (OFO) controllers have already been shown to work well in practice, with design and implementation tailored to specific applications, with parameters usually chosen on a case-by-case basis. In this paper, we propose a framework for tuning Online Feedback Optimization controllers, finding a trade-off between the tracking performance of the system and sampling time constraints. We validated the framework in a pressure controller in a centrifugal compressor, achieving up to 87\% times better tracking than the approach based on steady-state tuning. However, the tuning process in this work required multiple iterations. Thus, there is potential in using other methods, for instance based on surrogate optimization or safe learning algorithms, especially in safety-critical applications.

\balance

\bibliography{root}            
                                                   
\end{document}